\documentclass[a4paper,11pt]{amsart}
\usepackage{amssymb}
\begin{document}

\hyphenation{gra-vi-ta-tio-nal re-la-ti-vi-ty Gaus-sian
re-fe-ren-ce re-la-ti-ve gra-vi-ta-tion Schwarz-schild
ac-cor-dingly gra-vi-ta-tio-nal-ly re-la-ti-vi-stic pro-du-cing
de-ri-va-ti-ve ge-ne-ral ex-pli-citly des-cri-bed ma-the-ma-ti-cal
de-si-gnan-do-si coe-ren-za pro-blem gra-vi-ta-ting geo-de-sic
per-ga-mon cos-mo-lo-gi-cal gra-vity cor-res-pon-ding
de-fi-ni-tion phy-si-ka-li-schen ma-the-ma-ti-sches ge-ra-de
Sze-keres con-si-de-red tra-vel-ling ma-ni-fold re-fe-ren-ces
geo-me-tri-cal in-su-pe-rable sup-po-sedly at-tri-bu-table}

\title[All relativistic motions can be geodesically described]
{{\bf All relativistic motions\\can be geodesically described}}

\author[Angelo Loinger]{Angelo Loinger}
\address{A.L. -- Dipartimento di Fisica, Universit\`a di Milano, Via
Celoria, 16 - 20133 Milano (Italy)}
\author[Tiziana Marsico]{Tiziana Marsico}
\address{T.M. -- Liceo Classico ``G. Berchet'', Via della Commenda, 26 - 20122 Milano (Italy)}
\email{angelo.loinger@mi.infn.it} \email{martiz64@libero.it}

\vskip0.50cm

\begin{abstract}
A suitable choice of the four components of the metric tensor
which are at our discretion allows to represent geodesically also
the non-gravitational motions.
\end{abstract}

\maketitle


\normalsize

\vskip1.20cm \noindent \textbf{1.} INTRODUCTION
\par
You find in previous papers (Loinger, 2008) the demonstration that
the motions of point-masses which interact only gravitationally
are geodesic -- for instance, the motions of the elements of the
emblematic ``cloud of dust''.

\par In this Note we prove that the presence of non-gravitational
interactions -- \emph{e.g.}, electromagnetic interactions -- does
not forbid a geodesic description of the trajectories of the
``dust'' corpuscles.

\vskip0.50cm \noindent \textbf{2.} THE PROOF
\par First of all, we consider an electrically charged ``cloud of
dust''. The pertinent Einstein and Maxwell equations are ($c=G=1;
\, j,k,r \ldots = 0,1,2,3)$:

\begin{equation} \label{eq:one}
R^{jk} - \frac{1}{2} \, g^{jk} R + 8\pi \varrho \, v^{j}v^{k} +
8\pi E^{jk}=0 \quad ;
\end{equation}

\begin{equation} \label{eq:two}
-\sigma v^{j} + (4\pi)^{-1} F^{jk}_{\,\,\,\,\,:k} = 0 \quad ,
\end{equation}

\begin{equation} \label{eq:twoprime}
F_{jk,r} + F_{kr,j} + F_{rj,k} = 0 \tag{2'} \quad ;
\end{equation}

where: the colon and the comma denote, respectively, a covariant
and an ordinary derivative; $\varrho$ is the mass density of the
``dust'', $E^{jk}$ is the stress-energy-momentum tensor of the
electromagnetic field $F^{jk}$, $\sigma$ is the charge density.

\par As it is well known, the field equations (\ref{eq:one}) have
as a consequence:

\setcounter{equation}{2}
\begin{equation} \label{eq:three}
(\varrho \, v^{j}v^{k} + E^{jk})_{:k} = 0 \quad ,
\end{equation}

from which -- owing to equations (\ref{eq:two})
--(\ref{eq:twoprime}) and to the equations of matter conservation
$(\varrho v^{j})_{:j}=0$ -- we get the equations of motion of the
``dust'' elements:

\begin{equation} \label{eq:four}
\varrho \, v_{j:k} \, v^{k} + \sigma F_{jk} v^{k} = 0 \quad .
\end{equation}

This result does not depend on the choice of a particular
reference frame.

\par Now, let us choose the four components of the metric tensor
$g_{jk}$ which are at our discretion in such a way to satisfy the
four conditions:

\begin{equation} \label{eq:five}
E^{jk}_{\,\,\,\,\,:k} = 0 \quad ;
\end{equation}

then, if we call $g^{*}_{jk}$ the new components of the metric
tensor, we have, with an evident notation:

\begin{equation} \label{eq:six}
(\varrho \, v^{*j}v^{*k})_{:k} = v^{*j}_{\,\,\,\,:k} \, v^{*k}= 0
\quad ,
\end{equation}

\emph{i.e.}, a \textbf{\emph{geodesic}} motion. Remark that eqs.
(\ref{eq:six}) follow from Einstein equations (\ref{eq:one}),
\emph{they are independent of Maxwell equations} (\ref{eq:two}) --
(\ref{eq:twoprime}).

\par The above considerations can be immediately generalized to
the case of a material stress-energy-momentum tensor given by

\begin{equation} \label{eq:seven}
\varrho \, v^{j}v^{k} + E^{jk} + X^{jk} \quad ,
\end{equation}

where $X^{jk}$ is the sum of the matter tensors of further fields,
in particular of the field that characterizes a given continuous
medium.

\par Indeed, if we choose new $g^{*}_{jk}$'s in such a way that

\begin{equation} \label{eq:eight}
(E^{jk}+X^{jk})_{:k} = 0 \quad ,
\end{equation}

we see that the ``dust'' elements perform the
\textbf{\emph{geodesic}} trajectories given by eqs.
(\ref{eq:six}).

\par We remark finally that the choice of the two metric forms
corresponding, respectively,  to eqs. (\ref{eq:five}) and to eqs.
(\ref{eq:eight}) implies passages to two new Riemann-Einstein
manifolds.

\par (According to a famous similitude, we can imagine the
four-dimensional world as a mass of plasticine -- which can be
continuously deformed -- traversed by the world lines of the
material particles).

\vskip1.20cm \noindent \textbf{3.} A COROLLARY
\par The papers quoted in Loinger (2008) and the present paper
have a momentous consequence with regard to the gravitational
waves. A consequence which is quite clear to any unprejudiced
reader: the geodesic motions in general relativity are perfectly
analogous to the rectilinear and uniform motions in Maxwell theory
-- now, an electric charge in a rectilinear and uniform motion
does not emit electromagnetic waves.

Loinger, A. (2008). On gravitational motions. arXiv:0804.3991 v1
$[$physics.gen.ph$]$ 24 Apr 2008, 1-12. And the references
therein.

\end{document}